\documentclass{article}
\usepackage{spconf,amsmath,graphicx, hyperref}
\usepackage{xcolor}
\usepackage{tabularx}
\usepackage{makecell}
\usepackage{float}
\usepackage{multirow}
\usepackage{booktabs}
\usepackage{subfig}
\usepackage[square, comma, sort&compress,numbers]{natbib}
\usepackage[accsupp]{axessibility} 
\usepackage{flushend,soul}
\usepackage{enumitem}
\usepackage{microtype}
\newcommand\norm[1]{\left\lVert#1\right\rVert}
\newcommand{\red}[1]{\textbf{\textcolor{red}{#1}}}
\newcommand{\blue}[1]{\textit{\textcolor{blue}{#1}}}

\title{PocketDVDNet: Realtime Video Denoising for Real Camera Noise}

\name{Crispian Morris*, Imogen Dexter*, Fan Zhang, David R. Bull and Nantheera Anantrasirichai \thanks{The work was funded by the University of Bristol, the UKRI MyWorld Strength in Places Programme (SIPF00006/1), and the BBC via an iCASE scholarship. Real-world footage was provided by Humble Bee Films. *Equal contribution.}}
\address{Bristol Vision Institute, University of Bristol, One Cathedral Square, Bristol, BS1 5DD, UK.\\
{\{\href{mailto:crispian.morris@bristol.ac.uk}{crispian.morris}, \href{mailto:xs23823@bristol.ac.uk}{xs23823}, \href{mailto:fan.zhang@bristol.ac.uk}{fan.zhang}, \href{mailto:dave.bull@bristol.ac.uk}{dave.bull}, \href{mailto:n.anantrasirichai@bristol.ac.uk}{n.anantrasirichai}\}@bristol.ac.uk}
}

\begin{document}
\maketitle

%----------------------------------------------------------------------------------------
%	Abstract
%----------------------------------------------------------------------------------------
\begin{abstract}

Live video denoising under realistic, multi-component sensor noise remains challenging for applications such as autofocus, autonomous driving, and surveillance. We propose PocketDVDNet, a lightweight video denoiser developed using our model compression framework that combines sparsity-guided structured pruning, a physics-informed noise model, and knowledge distillation to achieve high-quality restoration with reduced resource demands. Starting from a reference model, we induce sparsity, apply targeted channel pruning, and retrain a teacher on realistic multi-component noise. The student network learns implicit noise handling, eliminating the need for explicit noise-map inputs. PocketDVDNet reduces the original model size by 74\% while improving denoising quality and processing 5-frame patches in real-time. These results demonstrate that aggressive compression, combined with domain-adapted distillation, can reconcile performance and efficiency for practical, real-time video denoising.

\end{abstract}

\maketitle

%----------------------------------------------------------------------------------------
%	Introduction
%----------------------------------------------------------------------------------------
\section{Introduction}
\label{sec:intro}

Noise in live video capture presents significant challenges for modern camera systems, particularly when focusing on moving subjects. In practice, sensor noise is rarely limited to a single source; shot noise, read noise, and structured or periodic artifacts combine to produce complex, scene-dependent degradations, especially pronounced at high ISO~\cite{Martinec2008NoiseDynamicRange}. This introduces a critical time constraint: unlike post-processing tasks, where extended computation time is acceptable, applications such as camera autofocus, autonomous vehicles, and video surveillance must operate within milliseconds to track moving objects effectively. Yet despite many recent advances, real-time video denoising under realistic noise conditions remains a challenge as yet unsolved.

\begin{figure}[ht]
    \centering
    \includegraphics[width=\linewidth]{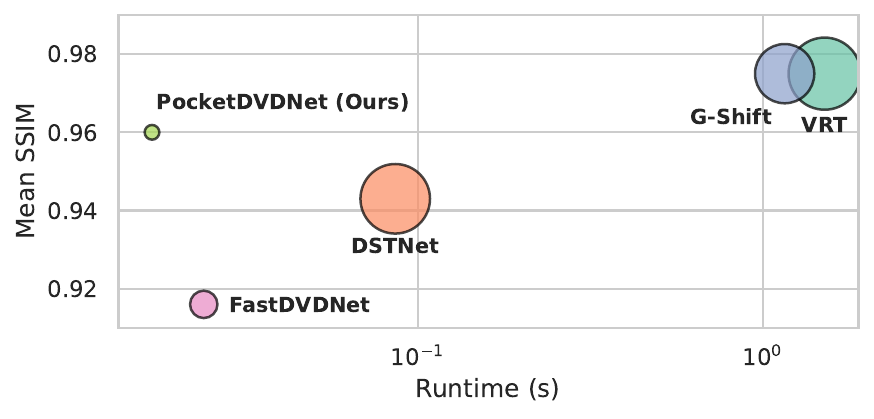}   
    \caption{Mean SSIM scores of all datasets from the models presented in~\autoref{tbl:quantitative_comp}. Our method performs extremely competitively given its size and runtime.}
    \label{fig:model_comparison}
    \vspace{-4mm}
\end{figure}

Traditional video denoising methods such as BM3D~\cite{bm3d2007} and VBM4D~\cite{vbm4d} rely on patch-based non-local filtering across spatial and temporal dimensions. While effective for moderate stationary noise, these classical approaches struggle with the more complicated characteristics of real video noise and are not useful for  real-time denoising. Deep learning has significantly advanced the state of the art in video denoising, and convolutional neural networks (CNNs) remain a popular approach for signal restoration and low-level vision tasks, due to their adaptability and suitability for implementation on portable devices. For example, DVDNet~\cite{tassano2019dvdnet} combines separate spatial and temporal processing networks for enhanced performance. Its lighter version, FastDVDNet~\cite{tassano2020fastdvdnet}, processes 5-frame temporal patches, achieving reasonable denoising quality while approaching real time, taking 0.024 seconds to denoise a 480p colour frame on a NVIDIA RTX 4090 GPU. However, it still falls short of the performance requirements for live denoising tasks like autofocus. Recent transformer approaches like VRT~\cite{liang2024vrt} deliver state-of-the-art performance but come at a resource cost, with 18.3 million parameters and heavy memory use. Shift-Net~\cite{li2023shiftnet} provides a better balance between performance and speed, offering variants trading accuracy and size.

%  mention in the above paragraph

This computational bottleneck has driven interest in neural network compression techniques. Methods have evolved from unstructured weight pruning~\cite{han2015deep} to more structured approaches that remove entire filters~\cite{sui2021chip}. Knowledge distillation~\cite{hinton2015distilling} has emerged as a complementary technique, enabling compact student networks to recover performance by learning from larger teachers. Theoretical advances like the Lottery Ticket Hypothesis~\cite{frankle2018lottery} have given insights into why aggressive compression succeeds, while methods like OBProxSG~\cite{chen2021orthant} have provided ways to induce sparsity in networks during training. Recent works like ST-MFNet Mini~\cite{morris2023st} and MTKD~\cite{jiang2024mtkd} demonstrate the effectiveness of combining pruning and distillation frameworks to video processing tasks.

A growing body of work also highlights the limitations of training denoisers on synthetic Gaussian noise. Zhou et al.~\cite{zhou2020awgn} show that models trained on additive white Gaussian noise (AWGN) generalise poorly to the complex, spatially correlated noise found in real images. Despite this, many video denoisers continue to rely on simplistic noise models, limiting their applicability in real-world conditions. This paper addresses these challenges through a model compression framework that reduces inference time and model complexity while \textit{improving} denoising efficacy. Our approach combines sparsity-guided pruning with knowledge distillation using a retrained large teacher model, and implements a realistic physics-informed noise model from Lin et al. \cite{lin2025generalpurposezeroshotsyntheticlowlight}, resulting in PocketDVDNet: a compressed architecture that achieves 34.86 dB while processing 5-frame temporal patches in real-time under diverse noise conditions. Figure \ref{fig:model_comparison} 1 demonstrates PocketDVDNet's favourable size, performance, and runtime against existing models.

Our main contributions are as follows:
\begin{itemize}[leftmargin=*, itemsep=-2pt]
    \item A workflow capable of creating domain-adapted real-time video denoising models, specifically for physics-informed real-world noise.
    \item We demonstrate that deep temporal networks can sustain aggressive compression without quality degradation.
    \item To our knowledge, the first real-time video denoiser that implements a realistic physics-informed noise model.
\end{itemize}

%----------------------------------------------------------------------------------------
%	Proposed Method
%----------------------------------------------------------------------------------------
\section{Proposed Method}
\label{sec:method}

\begin{figure}[tbp]
\centering
\includegraphics[width=\columnwidth]{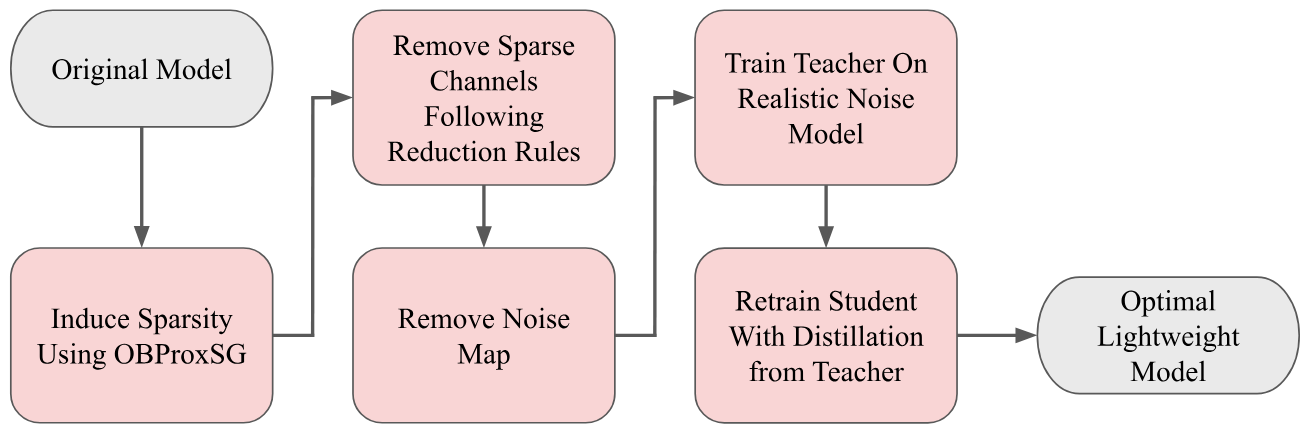}
\caption{The proposed workflow consists of model compression via sparsity-guided pruning followed by performance recovery through knowledge distillation}
\label{fig:flowchart}
\end{figure}

Our compression framework is illustrated in~\autoref{fig:flowchart}. It first employs OBProxSG~\cite{chen2021orthant} to induce sparsity in a given network, then uses the resulting sparse structure to guide targeted pruning and channel removal. Performance recovery employs two key modifications to restore denoising quality after compression. Knowledge distillation allows us to remove the explicit noise map inputs in the student since the teacher handles noise internally, simplifying the student architecture while enabling direct noisy-to-clean mapping. The teacher is retrained on a physics-informed noise model rather than simple Gaussian noise to better handle real-world noise conditions and improve denoising quality during distillation. 
This framework is applied to the FastDVDNet~\cite{tassano2020fastdvdnet} architecture, which is chosen for its excellent speed, but we hypothesize that this workflow would apply to any video restoration model.

\begin{figure}[tbp]
\centering
\includegraphics[width=\columnwidth]{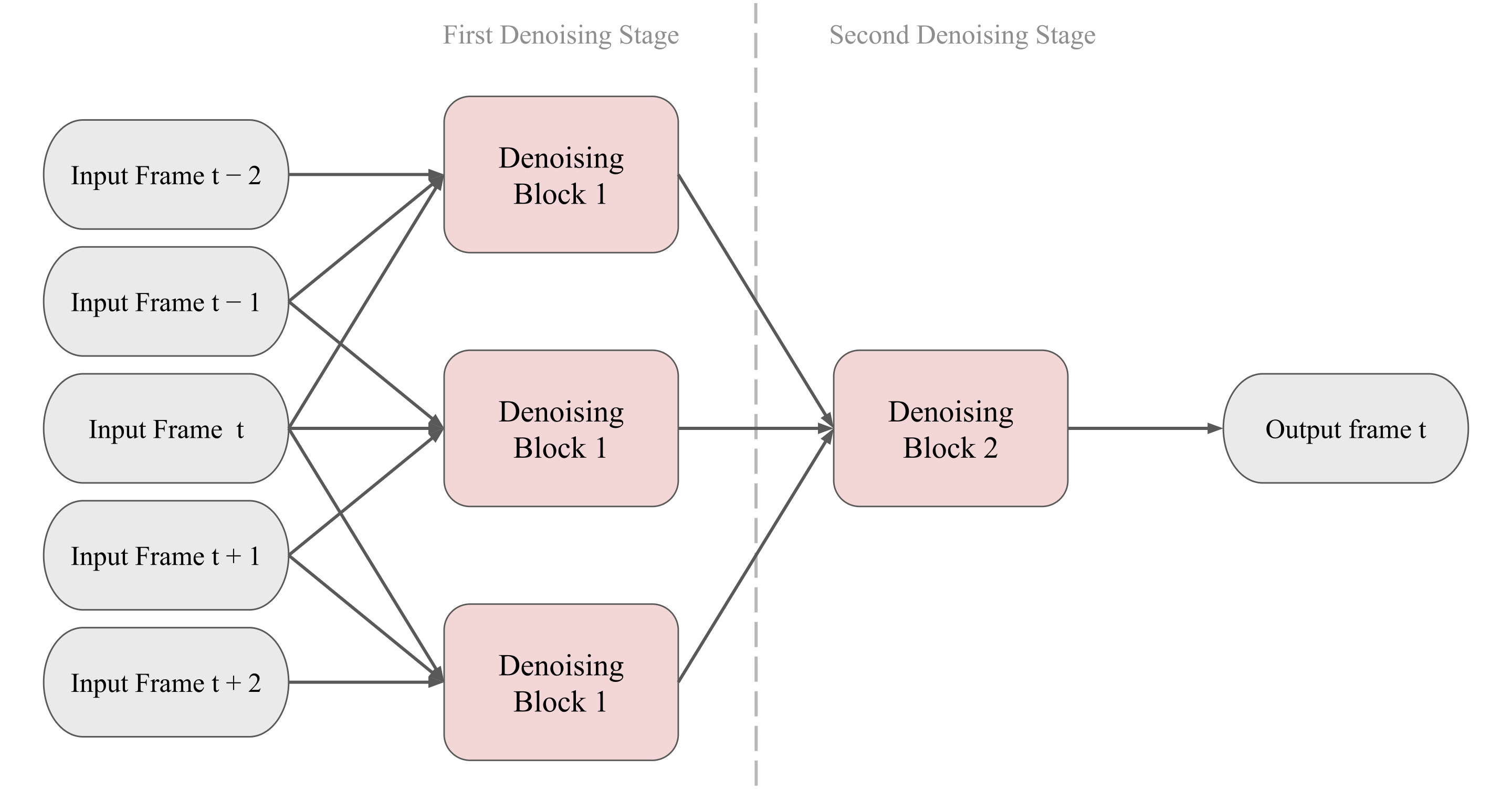}
\caption{PocketDVDNet processes 5-frame temporal patches through two cascaded denoising stages. }
\label{fig:fastdvd}
\end{figure}

% \begin{align}
% x'_{t-1} &= \mathcal{F}_{\text{temp1}}(x_{t-2},\, x_{t-1},\, x_{t},\, \sigma) \\
% x'_{t}   &= \mathcal{F}_{\text{temp1}}(x_{t-1},\, x_{t},\, x_{t+1},\, \sigma) \\
% x'_{t+1} &= \mathcal{F}_{\text{temp1}}(x_{t},\, x_{t+1},\, x_{t+2},\, \sigma)
% \end{align}

% followed by a second stage that processes the outputs:

% \begin{equation}
% \hat{x}_t = \mathcal{F}_{temp2}(x'_{t-1}, x'_t, x'_{t+1}, \sigma) 
% \end{equation}

% where $x_{t-2:t+2}$ represents the 5-frame input sequence, $\sigma$ is the noise standard deviation map, and $\hat{x}_t$ is the denoised central frame. Each denoising block follows a U-Net structure with grouped convolutions in the input layer:

% \begin{equation}
% \text{InputConv}: \mathbb{R}^{18 \times H \times W} \rightarrow \mathbb{R}^{32 \times H \times W}
% \end{equation}

% where the input concatenates 3 RGB frames with noise maps (6 channels per frame: 3 RGB + 3 noise map channels).
% % ABOVE MIGHT BE TOO LONG OR IN THE WRONG PLACE - COULD REPLACE WITH DIAGRAM

\begin{figure*}[ht]
    \centering
        % --- First row: plain images (no subfloat labels) ---
        \includegraphics[width=0.139\linewidth]{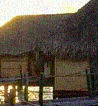}\;\!\!
        \includegraphics[width=0.139\linewidth]{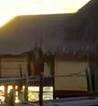}\;\!\!
        \includegraphics[width=0.139\linewidth]{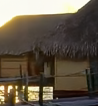}\;\!\!
        \includegraphics[width=0.139\linewidth]{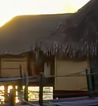}\;\!\!
        \includegraphics[width=0.139\linewidth]{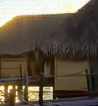}\;\!\!
        \includegraphics[width=0.139\linewidth]{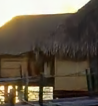}\;\!\!
        \includegraphics[width=0.139\linewidth]{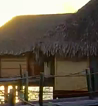}\\
        \vspace*{-0.35cm}
    
        % --- Second row: subfigures with labels ---
        \subfloat[Noisy]{\includegraphics[width=0.139\linewidth]{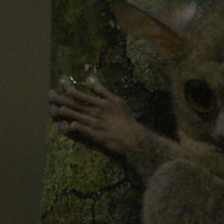}}\;\!\!
        \subfloat[DSTNet]{\includegraphics[width=0.139\linewidth]{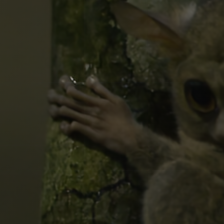}}\;\!\!
        \subfloat[VRT]{\includegraphics[width=0.139\linewidth]{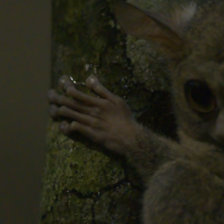}}\;\!\!
        \subfloat[G-Shift]{\includegraphics[width=0.139\linewidth]{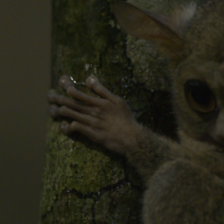}}\;\!\!
        \subfloat[FastDVDNet]{\includegraphics[width=0.139\linewidth]{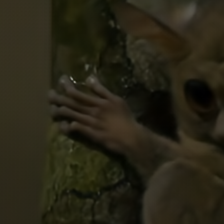}}\;\!\!
        \subfloat[Ours]{\includegraphics[width=0.139\linewidth]{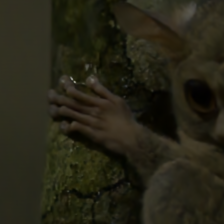}}\;\!\!
        \subfloat[GT]{\includegraphics[width=0.139\linewidth]{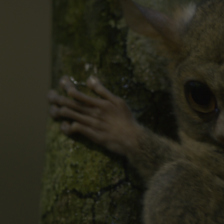}}\\[-0.9em]
    
        \vspace*{0.05cm}
        
        \caption{Qualitative examples demonstrating the superiority and shortcomings of our approach. Top: a sequence from the DAVIS test set, with synthetic noise. Bottom: a real-world example, with a 100 frame average as a pseudo GT. Zoom in for details.}
    \label{fig:qualitative}
    \vspace*{-0.2cm}
\end{figure*}

%----------------------------------------------------------------------------------------
\subsection{Model Pruning}
\label{sec:pruning}

We use OBProxSG~\cite{chen2021orthant} to induce sparsity during training, following \cite{morris2023st}. This allows the network to co-evolve its sparse structure alongside stable optimisation paths, a method leveraging lottery ticket research~\cite{frankle2018lottery}. The optimiser alternates between proximal gradient steps, which encourage sparsity, and orthant steps, whose primary role is to aggressively reinforce early sparsity patterns. However, in our pruning setting, orthant steps were found to introduce oscillation and destabilise training; therefore, only proximal steps were used.

The training objective $\mathcal{L}_{sparse}$ combines reconstruction loss with L1 regularisation
\begin{equation}
\mathcal{L}_{sparse} = \mathcal{L}_{Charb} + \lambda \norm{W}_1 .
\end{equation}
In our application, $\lambda$ uses warm-up scheduling to gradually introduce sparsity. The proximal gradient update applies soft thresholding $\text{prox}_{\lambda}(w) = \text{sign}(w) \cdot \max(|w| - \lambda, 0)$ to each weight. This shrinks small weights toward zero, inducing channel-wise sparsity patterns used to guide structured compression. Based on sparsity analysis, we calculate compression ratios for each layer $l$ as
\begin{equation}
r_l = \frac{\text{non-zero weights in layer } l}{\text{total weights in layer } l}
\end{equation}

% Channel reduction then follows $C_{new} = \text{reduction\_rule}(r_l \cdot C_{original})$ following these rules:

% \begin{itemize}
%     \item \textbf{Power-of-2:} Channels reduced to the nearest $\{16,32,64,128\}$
%     \item \textbf{Bottleneck avoidance:} Important pathway layers (input/output) preserved or minimally reduced
%     \item \textbf{Architectural constraints:} Final classification layers (3 channels) unchanged, and skip connection compatibility maintained
% \end{itemize}

Channel reduction, defined by $C_{new} = \text{reduction}(r_l \cdot C_{original})$, is governed by several key principles. Our \textbf{power-of-2} rule reduces channels to the nearest power of two, and \textbf{bottleneck avoidance} is practiced to ensure model integrity, preserving important input and output layers or reducing them minimally. Finally, we adhere to \textbf{architectural constraints}, which means that the final three-channel classification layers remain unchanged, and compatibility with skip connections is maintained.

The compressed model is then retrained, achieving 31.17dB PSNR on the DAVIS validation dataset~\cite{perazzi2016benchmark} with $\sigma=50$ compared with FastDVDNet's 31.86dB baseline performance. This compression method reduces FastDVDNet's 2,482,336 parameters down to 648,096, a $74\%$ size reduction, while maintaining competitive performance. The resulting architecture is shown in Fig.~\ref{fig:fastdvd}.

\subsection{Architecture Modifications}
\label{sec:frames}

% Aiming to mitigate the performance loss from compression, \hl{we tested extending the input sequence from 5 to 7 frames. However, in experiments, this temporal extension did not show an improvement over the baseline, so we kept the 5-frame architecture.}

We remove the explicit noise map inputs that FastDVDNet concatenates with each frame. While such maps can be beneficial for high-capacity models,  because they are inherently stochastic, they may introduce additional variance during training. We propose they could act as a 'distraction' for a lightweight student with limited representational capacity. By eliminating noise maps, the student learns to handle noise implicitly through distillation, optimizing a more direct and stable mapping from noisy inputs to clean outputs. This eliminates the need for explicit noise level inputs.

%----------------------------------------------------------------------------------------
\subsection{Multi-Component Noise Model}
\label{noise}

To further tailor our model to real-world capture, we adopt the recent physics-based noise model from Lin et al.~\cite{lin2025generalpurposezeroshotsyntheticlowlight}, which provides realistic sRGB noise synthesis. Unlike traditional approaches that train denoisers on simple additive Gaussian noise, real video capture contains complex, multi-component noise patterns. Our compressed architecture benefits from exposure to this full spectrum of noise to prepare it for practical applications.

The five-component noise model simulates distinct physical noise phenomena.~\textbf{Heteroscedastic noise} combines signal-dependent shot noise with signal-independent read noise.~\textbf{Quantisation noise} models artifacts from analogue-to-digital conversion using a uniform distribution. We also include \textbf{banding noise} to capture horizontal or vertical patterns commonly seen at high ISO settings, with \textbf{temporal banding} ensuring that these patterns remain consistent across video frames. Finally, \textbf{periodic noise} simulates electrical interference by introducing patterns in the frequency domain.

During each training iteration, parameters 
$\sigma_s, \sigma_r, \lambda_q, \sigma_b,$ \linebreak
$\sigma_{bt}, \sigma_{p1}, \sigma_{p2}, \sigma_{p3}$ 
control the intensity of each noise component and are randomly sampled in accordance with the values specified in \cite{lin2025generalpurposezeroshotsyntheticlowlight}. For our 5-frame temporal architecture, we maintain noise parameter consistency across each video patch while allowing spatial variation, to better model real camera noise.
Using this model forces the compressed network to learn robust representations that generalise across noise types, and also provides data augmentation that effectively increases training set diversity.

%----------------------------------------------------------------------------------------
\subsection{Knowledge Distillation}
\label{sec:distillation}

To further improve performance after our substantial network compression, we use knowledge distillation from a large state-of-the-art teacher, ShiftNet ~\cite{li2023shiftnet}, which is retrained using our multi-component noise model on the same dataset. ShiftNet provides excellent denoising performance through its grouped spatial-temporal shift operations, making it an effective teacher for our lightweight student, PocketDVDNet. Multi-teacher distillation was considered with VRT~\cite{liang2024vrt}, but due to memory constraints, this was not possible.

The overall distillation loss is a weighted combination of two terms:
\begin{equation}
\mathcal{L}_{total} = \alpha\mathcal{L}_{teacher} + (1-\alpha)\mathcal{L}_{gt},
\end{equation}
where $\alpha$ controls the balance between teacher guidance and ground truth supervision. We use $\alpha= 0.5$ to give equal weight to both learning objectives. For both loss terms, we employ the Charbonnier loss function, which provides better gradient properties than MSE for image tasks~\cite{Charbonnier1997DeterministicER}:  
\begin{equation}
\mathcal{L}_{teacher} = \frac{1}{N}\sum_{i=1}^{N}\sqrt{(y_{student}^i - y_{teacher}^i)^2 + \epsilon^2},
\end{equation}
\vspace{-2mm}
\begin{equation}
\mathcal{L}_{gt} = \frac{1}{N}\sum_{i=1}^{N}\sqrt{(y_{student}^i - y_{gt}^i)^2 + \epsilon^2},
\end{equation}
%\vspace{-3mm}
%
where $y_{student}$, $y_{teacher}$, and $y_{gt}$ represent the outputs of the student model, teacher model, and ground truth, respectively, and $\epsilon = 0.0001$. 

%----------------------------------------------------------------------------------------

\section{Results and Discussion}
\label{sec:results}

\subsection{Experiment settings}
\label{sec:experiments}

Our models are implemented in PyTorch and trained on a NVIDIA RTX 4090 GPU. For the pruning phase, we use learning rate $1.5 \times 10^{-5}$, batch size of 64, and the pruning regularisation parameter $\lambda_{reg} = 0.09$. For knowledge distillation, we use batch size 32 and learning rate $1 \times 10^{-3}$. Training is performed on video sequences from the DAVIS~\cite{perazzi2016benchmark} training set and the BVI-AOM dataset~\cite {nawala2024bvi}. Each training sample consists of clean video frames corrupted with realistic noise using the model and the random values employed by Lin et al.~\cite{lin2025generalpurposezeroshotsyntheticlowlight}. 
We assess denoising quality using PSNR and SSIM, and efficiency in speed (s) and parameter count.

% Maybe mention running in real time on a live camera?

\begin{table}[t]
\caption{Quantitative comparison results (PSNR/SSIM) for PocketDVDNet and other methods. For each column, the best result is bold in red and the second best is italicised in blue.}
\vspace{-3mm}
	\begin{center}
		\resizebox{\linewidth}{!}{
			\begin{tabular}{lcccr}
				\toprule
                Model & Set8 & DAVIS &  RT & \#P \\
                \midrule 
                VRT~\cite{liang2024vrt}                     &  \red{36.704}/\red{0.971}  &  \red{38.605}/\blue{0.979}   &  1.505  & 18.3 \\
                DSTNet~\cite{pan2023deep}                   &  32.090/\blue{0.938} &  33.967/0.947  &  0.086  & 17.0 \\
                % RVRT~\cite{liang2022recurrent}              &  / &  /   &  0.  & 13.6 \\
               % Other Model?~\cite{tassano2020fastdvdnet}   &  / &  /  &  /  &  0.  & 29.23 \\
                G-Shift Net~\cite{li2023shiftnet}           &  \blue{36.639}/\red{0.971} &  \blue{38.432}/\blue{0.979} &  1.153  & 12.3 \\
                FastDVDNet~\cite{tassano2020fastdvdnet}     &  32.493/0.924 &  32.996/0.908  &   \blue{0.024}  & \blue{2.5} \\
                \midrule
                %PocketDVDNet (7 Frames) & 31.778,0.917 & 33.616/0.921  & / & 0.017 & \red{0.6} \\ 
                PocketDVDNet & 32.987/0.933 &  34.857/\red{0.986} & \red{0.017} & \red{0.6} \\

				\bottomrule
		\end{tabular} }
		\label{tbl:quantitative_comp}
  		\vspace{-9mm}
	\end{center}
\end{table}
\subsection{Quantitative Comparison}

% Present results clearly. Use tables for quantitative data and figures for qualitative comparisons. More figures in this section than words is fine.

Our method is compared against various state-of-the-art models in \autoref{tbl:quantitative_comp}. DSTNet Plus~\cite{pan2025learning}, which has a specific denoising variant, is not compared against due to a lack of code, and instead, the original DSTNet is trained to denoise using our approach. Models are re-trained on the same realistic noise model~\cite{lin2025generalpurposezeroshotsyntheticlowlight} as ours for a fair comparison. The average runtime (RT) in seconds for denoising a 480p DAVIS frame, as well as the number of model parameters (\#P) in millions for each method, are also reported. PocketDVDNet notably outperforms its teacher, appearing to improve generalization under realistic noise conditions. This supports the hypothesis that the removal of the noise map better allows the student to optimize.

%----------------------------------------------------------------------------------------
\subsection{Qualitative Comparison}

Examples of denoised frames processed by the tested models are shown in~\autoref{fig:qualitative}. The proposed model’s performance is observably closer to the ground truth than that of FastDVDNet, which fails to capture the complexity of the superior noise model. However, in low-light sequences, our model fails to recover as much detail as the larger models. This will be a subject of our future work.

%----------------------------------------------------------------------------------------
\subsection{Ablation Studies}
\label{sec:ablation}

We validate each component's contribution on the DAVIS~\cite{perazzi2016benchmark} validation dataset in Table \ref{tab:ablation}. Pruning and AWGN experiments use $\sigma = 50$ noise, while experiments on the physics-based noise model use 100 randomly sampled real camera noise parameters applied consistently across frames. PSNR, parameters (\% of baseline), and 480p runtime (s) are reported.
\begin{table}
\centering
\caption{Ablation study on distillation, noise map removal, and tested on a portable chip.}

\label{tab:ablation}
\begin{tabular}{lccr}
\toprule
Method & PSNR (dB) & \#P ($\%$) & RT(s) \\
\midrule
Baseline FastDVDNet & 31.86 & 100\% & 0.024 \\
Pruned FastDVDNet & 31.17 & 26\% & \ 0.019 \\
\midrule
\textit{Distillation:} & & & \\
AWGN & 31.34 & 26\% & 0.019 \\
Realistic noise & 30.83 & 26\% & 0.019 \\
Realistic w/o noise map & \textbf{34.86} & 26\% & \textbf{0.017}
\\
\midrule
% {PocketDVDNet:} & & & \\
Ours w/ Apple M3 Pro & 34.06 & 26\% & 0.105 \\
\bottomrule
\end{tabular}
\vspace{-2mm}

\end{table}

%----------------------------------------------------------------------------------------
%	Conclusion
%----------------------------------------------------------------------------------------
\vspace{-2mm}
\section{Conclusion}
\label{sec:conclusion}

We present PocketDVDNet, a lightweight model capable of real-time inference on realistic noise, addressing a limitation that has prevented learning-based denoisers in time-critical applications like autofocus and live video. 
Our compression and recovery framework demonstrates that aggressive parameter reduction can be accompanied by improved performance. Despite having 74\% fewer parameters than FastDVDNet, PocketDVDNet delivers higher PSNR and runs 30\% faster. 
The success in our performance recovery stems from two key components: i) retraining the teacher model on realistic noise enables effective domain transfer to real-world conditions, and ii) knowledge distillation allows the complete removal of explicit noise map inputs from the student architecture. The compressed network learns implicit noise handling from the teacher, outperforming its own limited noise modeling. While demonstrated on FastDVDNet, our approach provides a general template for compressing temporal models for deployment, including on mobile devices. Future research could explore extending this framework to transformer-based architectures or other low-level vision processing tasks.

%------------------------------------------------------------------- -------------------
%	References---------------------------------------------------------------------------
\small
\bibliographystyle{ACM-Reference-Format}
\bibliography{bibliography.bib} 

\end{document}